\documentclass[smallextended,natbib]{svjour3}
\smartqed  
\usepackage{pifont}
\usepackage{amsmath}
\usepackage[caption=false]{subfig}
\usepackage{multirow}
\usepackage{soul}
\usepackage{array}
\usepackage{verbatim}
\usepackage{threeparttable}
\usepackage{tabularx}
\newcolumntype{Y}{>{\centering\arraybackslash}X}
\usepackage{enumitem}
\usepackage{makecell}
\usepackage{tabu}
\setlist{noitemsep}
\usepackage{url}
\usepackage[textsize=scriptsize,backgroundcolor=green]{todonotes}
\usepackage[left=0.85in,right=0.85in,top=0.85in,bottom=0.75in,%
            footskip=.25in]{geometry}

\newcommand{\evetar}{\emph{EveTAR}}
\newcommand{\sevetar}{E\MakeLowercase{ve}TAR}
\definecolor{lightblue}{rgb}{.50,.95,1}
\definecolor{tri}{rgb}{.25,.88,.82}
\definecolor{lilac}{rgb}{0.85,0.64,0.85}

\newcommand{\hlc}[2][yellow]{ {\sethlcolor{#1} \hl{#2}} }
\newcommand\mh[1]{\hlc[lilac]{{\bf MH}: #1}}

\newcommand\mk[1]{\hlc[green]{{\bf MK}: #1}}
\newcommand\rs[1]{\hlc[yellow]{{\bf RS}: #1}}

\newcolumntype{?}[1]{!{\vrule width #1}}

\begin{document}
\title{\sevetar: Building a Large-Scale Multi-Task Test Collection over Arabic Tweets\thanks{This manuscript describes a major extension to an earlier preliminary version published at SIGIR'16~\citep{almerekhi2016evetar}. Improvements over the preliminary work include providing a much deeper justification of the design choices made during the creation of the collection, extension of the test collection to support two additional tasks, improvements to the judgments collected, providing four subsets of the collection to increase its accessibility and use case scenarios, and running experiments that demonstrate the reliability of the proposed test collection. Improvements to the judgments include filtering out topics with low-agreement among annotators, removal of inaccessible tweets from the document collection, collecting additional relevance judgments to increase the qrels set size, and collecting novelty judgments to allow the collection to support two tasks (timeline generation and real-time summarization).}}
\author{Maram Hasanain\and Reem Suwaileh\and Tamer Elsayed\and Mucahid Kutlu\and Hind Almerekhi}


\institute{Maram Hasanain, Reem Suwaileh, Tamer Elsayed, Mucahid Kutlu \at
              Department of Computer Science and Engineering,
College of Engineering, Qatar University, Qatar\\
              \email{maram.hasanain,reem.suwaileh,telsayed,mucahidkutlu@qu.edu.qa}           
           \and
           Hind Almerekhi \at
					Hamad Bin Khalifa University, Qatar
					\\
              \email{halmerekhi@qf.org.qa}         
}

\date{Received: date / Accepted: date}

\maketitle

\begin{abstract}

This article introduces a new language-independent approach for creating a large-scale high-quality test collection of tweets that supports multiple information retrieval (IR) tasks without running a shared-task campaign. The adopted approach (demonstrated over Arabic tweets) designs the collection around \emph{significant} (\emph{i.e., popular}) \emph{events}, which enables the development of topics that represent frequent information needs of Twitter users for which rich content exists. That inherently facilitates the support of multiple tasks that generally revolve around events, namely event detection, ad-hoc search, timeline generation, and real-time summarization. The key highlights of the approach include diversifying the judgment pool via interactive search and multiple manually-crafted queries per topic, collecting high-quality annotations via 
crowd-workers for relevancy and in-house annotators for novelty, filtering out low-agreement topics and inaccessible tweets, 
and providing multiple subsets of the collection for better availability. Applying our methodology on Arabic tweets resulted in \evetar{}, the first freely-available tweet test collection for multiple IR tasks. \evetar{} includes a crawl of 355M Arabic tweets and covers 50 significant events for which about 62K tweets were judged with substantial average inter-annotator agreement (Kappa value of 0.71). We demonstrate the usability of \evetar{} by evaluating existing algorithms in the respective tasks. Results indicate that the new collection can support reliable ranking of IR systems that is comparable to similar TREC collections, while providing strong baseline results for future studies over Arabic tweets.

\end{abstract}

\maketitle

\section{Introduction} 
\label{intro}
Since the start of the 
Arab Spring in 2011, the Arab world has faced a continuous stream of events that changed the way Arab people are using online micro-blogging services such as Twitter.
Several key events were initiated over those platforms and real-time updates were 
more frequent there than in typical mainstream media~\citep{bruns_arab_2013,wasike_framing_2013}.
With an average of 19.7M Arabic tweets posted every day
as of March 2016~\citep{smr17}, the Arab users became more interested in several online tasks that are related to events. Among those tasks, they might be interested in automatically detecting events as soon as they occur~\citep{alsaedi2016sensing} (\textit{Event Detection}), or once an event occurs, they become interested in searching for tweets that are discussing that event~\citep{darwish2012language} (\textit{Ad-hoc Search}), or getting real-time updates about the events as they are developing over time~\citep{Magdy2016513} (\textit{Real-time Summarization}), or even asking for a timeline of the event when it concludes
(\textit{Timeline Generation}).
That makes the need for automatic tools that can effectively search huge tweet stream, detect events before being publicly announced, and summarize their timelines obvious and imperative. 
 
To design and develop high-quality Information Retrieval (IR) systems over Twitter, evaluation mechanisms such as \emph{test collections} are evidently required. A test collection typically consists of a set of documents (e.g., tweets), topics representing information needs (e.g., events), and relevance judgments (i.e., labels) that specify which documents are relevant to the topics~\citep{sanderson2010}.
There are several publicly-available test collections over Twitter that support a broad range of retrieval tasks~\citep{ounis2011overview,soboroff2012overview,lin2013overview,lin2014overview,lin2015overview,lin2016overview} 
and hence motivated a large body of research; however none of those collections is in Arabic, which severely hinders the advancement of research on Arabic tweets.
Moreover, the majority of those collections were constructed using shared-task evaluation campaigns, which benefit from the participation of multiple research teams that contribute to the pool of judged documents, in addition to the employment of experienced dedicated annotators; that is costly and not easily feasible in many languages of relatively-low IR research resources such as Arabic.

In this article, we address the problem of building a large-scale high-quality test collection of tweets that supports multiple IR tasks without running a shared-task campaign. To tackle the problem, we adopted a new language-independent approach that designs the collection around \emph{significant} events. A significant event is defined as an occurrence that happens at a particular time in a specific location and is discussed by the media (e.g., covered by an online news article). Focusing on significant events has several advantages. First, it enables the development of topics that represent frequent information needs of Twitter users as they often search Twitter for timely and up to date information about events~\citep{teevan2011twittersearch, zhao2013questions}. Second, rich content exists for popular events. Third, the popularity of the topics might make the annotations much easier and thus consistent.
 Finally, it inherently facilitates the support of multiple tasks that generally revolve around events.

Applying our approach on \emph{Arabic} tweets resulted in \textbf{\evetar{}}, our \textbf{Eve}nt-centric \textbf{T}est Collection of \textbf{Ar}abic Tweets\footnote{pronounced as evi-tar.}. \evetar{} is built over a one-month crawl of 355M Arabic tweets and supports four IR tasks: event detection, ad-hoc search, timeline generation, and real-time summarization.
The topics cover 50 events in that month selected using Wikipedia's Current Events Portal\footnote{\url{https://en.wikipedia.org/wiki/Portal:Current_events}}. A diversified judgment pool was constructed via multiple queries per topic that are manually-crafted using interactive search. To ensure large-scale but high quality judgments, we first employed 
crowd-workers to collect relevance judgments, then we filtered out the topics with the lowest inter-annotator agreement and removed inaccessible tweets, reaching an average  \emph{substantial} agreement indicated by Kappa value of 0.71 over 62K judgments. As two of the supported tasks require the identification of novel tweets per topic, we finally added novelty annotations via in-house annotators to ensure the integrity of the annotations. 

We demonstrate the usability of \evetar{} by evaluating a number 
of strong existing techniques in each of the supported tasks. The performance results of those systems constitute reference baselines for future studies in the respective research problems. Moreover, we show that ranking those systems over \evetar{} is at least as reliable as in comparable TREC collections.

With the huge size of the collection and the sharing restrictions mandated by Twitter, we have provided a smaller sample (about 15M tweets) of the collection that is much easier to crawl and showed that it is as reliable as the full collection in ranking the tested systems. Furthermore, two subsets of that sample that include only Modern Standard Arabic (MSA) tweets and only dialectal tweets are also provided to encourage IR research on Arabic dialects, which naturally exhibit different unique characteristics from MSA and are widely used by Arab users in social media platforms~\citep{darwish2012language,darwish2014arabic}. Performance results of the evaluated systems are presented on all subsets of \evetar{} as well as the full set (whenever possible). 

Our contribution in this work is 5-fold:
\begin{enumerate}
	\item We propose a language-independent approach to build a multi-task test collection without the need of a shared-task campaign.
    \item We introduce \evetar{}, the \textit{first} large-scale test collection over \textit{Arabic} tweets that supports event detection, ad-hoc search, timeline generation, and real-time summarization IR tasks. 
	\item We make the full test collection publicly-available for research\footnote{\url{http://qufaculty.qu.edu.qa/telsayed/evetar}}, including ids of 355M Arabic tweets, 50 events and 62K relevance judgments, novelty annotations, inter-annotator agreements, queries used to identify potentially-relevant tweets for the events, and documented design of the crowdsourcing tasks. We also provide the annotations per tweet to further support relevant studies on crowdsourcing in IR.
	\item We introduce four versions of \evetar{} that make accessibility of the collection easier and encourage research over MSA vs. dialectal tweets.
    \item We demonstrate the usability of \evetar{} by evaluating existing techniques over the four supported tasks. The resulting performance scores constitute reference baselines for future studies.
\end{enumerate}

The remainder of the article is organized as follows. We first discuss the related studies on building Arabic and tweet test collections in Section~\ref{rw}. We present our proposed approach in Section~\ref{approach}, and then we show how it is applied in practice on Arabic tweets (resulting in \evetar{}) in Section~\ref{buildcoll}. 
Section~\ref{eval} covers the evaluation of \evetar{} from usability and reliability perspectives. Finally, Section~\ref{sec:conclusion} concludes our study and gives some guidelines for future work.
\section{Related Work}\label{rw} 
As our test collection is over Arabic tweets, in this section we review existing Arabic test collections 
and survey literature on constructing tweet test collections.  
\subsection{Arabic Test Collections} \label{section_arabic_test_collections}


Test collections supporting Arabic IR can be categorized into two groups: multi-lingual collections and Arabic-only collections, and both are discussed next. For further information about Arabic test collections and Arabic IR, please refer to 
\citet{darwish2014arabic}. 
\subsubsection{Multi-lingual Test Collections} Many evaluation campaigns used Arabic documents in  multi-lingual test collections for various IR tasks such as cross-language retrieval ~\citep{gey2001trec,oard2002trec}, topic detection and tracking~\citep{strassel2005tdt4,glenn2006tdt5}, identifying nuggets (i.e., text snippets)\footnote{\url{https://www.ldc.upenn.edu/collaborations/current-projects/bolt}}, thematic grouping of retrieved results\footnote{\url{http://www.darpa.mil/program/broad-operational-language-translation}}, question answering for machine reading~\citep{clef2012QA}, and community question answering~\citep{alessandromoschitti2015semeval,nakov2016semeval,SE_2017_task3}. These test collections focus on genres other than tweets in various languages including Arabic, while \evetar{} focuses only on Arabic tweets. 
\subsubsection{Arabic-only Test Collections} To the best of our knowledge, there is no prior work that constructs an Arabic-only test collection as a primary contribution.
Therefore, researchers working on Arabic IR systems had to construct their own test collections to conduct their experiments. Examples of such collections are presented in this section.

\citet{darwish2012language} explored issues making Arabic microblog search harder and investigated solutions for some of those issues. In order to perform their experiments, they constructed an Arabic tweet test collection for ad-hoc search  
that contains 112M tweets, 35 topics and 19,824 relevance judgments. 
For each topic, they manually judged top 30 results returned by each of 7 IR systems they employed in their experiments. 
In comparison, \evetar{} is larger in all aspects. To detect potentially-relevant documents, we used multiple queries per topic instead of multiple IR systems. Moreover, \evetar{} is designed to evaluate several IR tasks including ad-hoc search.

\citet{alsaedi2015arabic} proposed an Arabic event detection method and labeled the output of their system for evaluation.
Experiments were conducted over 1M Arabic tweets restricted to Abu Dhabi. 3 annotators were hired to label the system output represented by event clusters. 
Differently, \evetar{} supports multiple IR tasks, not only event detection, and is more reusable because events in \evetar{} are not output of a specific system. In addition, events in \evetar{} are not restricted to a specific location, which helps increase the linguistic diversity of the collection and the chance of having different types of events. 

\citet{azmi_2016} developed a question answering test collection to evaluate a system for answering why-questions. Given 700 Arabic articles from different genres, they recruited an annotator to develop 100 Arabic why-questions and answers. The annotator created that set after browsing the articles to ensure that the developed questions can be answered from the collection. The system was finally evaluated by manually comparing the system answers to the human-provided answers. \evetar{} allows automatic evaluation for different IR tasks and targeting evaluation over Arabic tweets.
Overall, it is very challenging for researchers to work on today's IR research problems over Arabic data by employing existing Arabic test collections. 
The collections are usually  old and contain limited number of documents  and judgments. 
To the best of our knowledge, there is no publicly available test collection of Arabic tweets that serves the evaluation of event detection, real-time summarization, ad-hoc search and timeline generation tasks.

\subsection{Tweet Test Collections}  \label{section_tweet_test_collections}
Twitter 
provides a rich resource of information but the huge size of the stream invokes the need for IR systems to help users sift through the stream. However, available tweets test collections for IR system evaluation are limited compared to other document types. This can be due to tweets redistribution policy of Twitter
(i.e., redistribution of tweet texts at a scale larger than few tens of thousands is prohibited) and also being a more recent source compared to others such as Web pages and news articles. Existing tweets test collections were constructed in two ways: shared-task evaluation campaigns that develop collections to evaluate systems from teams participating in the same IR task, and centralized efforts by individual research teams who created test collections to develop and evaluate their IR systems or for the sake of providing test collections for the community. In this section, we present examples under each and also discuss what distinguishes \evetar{} from available collections.
\subsubsection{Shared-Task Evaluation Campaigns}
One of the leading efforts to develop tweet test collections was carried by TREC evaluation campaigns.
TREC developed large-scale tweet test collections as part of its Microblog track which ran from 2011 to 2015 and Real-time Summarization track which ran in 2016. The tracks targeted various IR tasks over English tweets such as ad-hoc search (2011-2014), filtering (2012), tweet timeline generation (2014) real-time filtering (2015), and real-time summarization (2016). 4 different collections were produced: 
\begin{itemize}
\item Tweets2011 corpus, which contains 16M tweets 
have been used in 2011~\citep{ounis2011overview} and 2012 tracks~\citep{soboroff2012overview}. TREC organizers employed 
\emph{download-tweets-yourself} approach, in which each participant has to download the tweets from twitter.com site given the ids of the 16M tweets in the collection. A total of 115 topics and 133K relevance judgments were created along that collection.
\item Tweets2013 corpus,  which contains 243M tweets 
has been used in 2013~\citep{lin2013overview} and 2014~\citep{lin2014overview} tracks. Differently from Tweets2011, in order to distribute the tweet collection, TREC employed evaluation-as-a-service paradigm, in which tweets are stored centrally and the participants can access the tweets via an online search API. In total, 108 topics and 129K relevance judgments are released with that corpus.
\item In 2015, instead of distributing the collection or providing online access to it ahead of time, each participant obtained the tweets directly from Twitter for a defined 10-day evaluation period~\citep{lin2015overview}. The participants in the track were responsible for collecting the 1\% random sample of tweets made freely-available by Twitter daily\footnote{\url{https://dev.twitter.com/streaming/reference/get/statuses/sample}}. The test collection included 51 topics and 94K relevance judgments in addition to novelty judgments. 
\item In 2016, TREC introduced the real-time summarization track~\citep{lin2016overview}, which also requires collecting tweets posted during a predefined evaluation period, instead of providing pre-crawled tweets. 51 topics and 67K relevance judgments in addition to novelty judgments where released with the collection.
\end{itemize}

Another evaluation campaign that targeted IR tasks over tweets is the FIRE 2016 Microblog track~\citep{ghosh2016overview} which focused on retrieval of tweets during disaster events. Organizers of the campaign collected tweets for the two weeks following the earthquake that occurred in Nepal and parts of India on the 25th of April 2015, by tracking the keyword ``nepal". After eliminating duplicate and near-duplicate tweets, their dataset contained 50K English tweets. They developed 7 topics for the evaluation. In contrast, \evetar{} contains much more tweets, and topics cover multiple independent events, while their collection focused on sub-events under a single large-scale event.

To the best of our knowledge, there is no prior work on building multi-task collections except in TREC 2012 and 2014 Microblog tracks. Each track included a test collection to evaluate two tasks: ad-hoc search and filtering in 2012~\citep{soboroff2012overview}, and ad-hoc search and timeline generation in 2014~\citep{lin2014overview} . 
Differently from those collections, \evetar{} focuses on Arabic tweets while TREC collections focused on English only. More importantly, \evetar{} is built without running a shared-task evaluation campaign where a diversified judgment pool was constructed using multiple manually-crafted queries per topic. Additionally, \evetar{} supports the event detection task while it has never been targeted by any TREC tracks related to tweets. Evaluating multiple systems per task using \evetar{} and TREC tweets test collections showed that \evetar{} is as reliable as TREC collections in ranking the tested systems (Section~\ref{section_reliability}). We provide an extensive comparison covering TREC test collections and \evetar{} in Section~\ref{section_putting_in_context}.
\subsubsection{Centralized Efforts}
Many studies proposing IR systems for tweets also build their own test collections but for the purpose of developing and evaluating their systems. Collecting judgments for the test collection is generally done by evaluating the output of the proposed systems, reducing the reusability of the labeled data. 
 Some of the tasks targeted by those studies include aligning tweets with events~\citep{rowe2012aligning}, tagging tweets~\citep{ma2014tagging}, event monitoring~\citep{zhang2015sense}, event detection~\citep{liu2016reuters,kunneman2014event,becker_2012,qin2013feature}, adaptive filtering~\citep{Magdy2016513} and first story detection~\citep{petrovic2012using}. The test collections used in the cited event-detection studies are smaller than \evetar{} since the main focus of those studies was to develop IR systems instead of building test collections.  

There are also studies mainly focusing on constructing English test collections. 
\citet{mcminn2013building} built a publicly-available test collection for evaluating event detection systems over English tweets. The authors crawled around 120M tweets, covering more than 500 events identified using automatic and manual ways, and collected labels for over 150K tweets via crowdsourcing.  
\evetar{} is designed for multiple tasks, not only for event detection. Additionally, we crawled about 3 times more tweets (355M vs. 120M) and labeled more tweets per event on average. CREDBANK is another test collection developed without running an evaluation campaign~\citep{mitra2015credbank}. The collection was designed to evaluate tasks focusing on information credibility on social media and  includes more than 60M tweets and 1049 events with crowd-sourced judgments. But we target different IR tasks. 

\section{Our Approach}
\label{approach}
A typical test collection comprises three major components: a document collection, a set of topics, and a set of judgments per topic. There are many challenges in test collection construction process. First, the document collection should be a good representative of the domain in which IR systems will be applied. Second, the topics should be carefully designed to represent real-world information needs. Third, the document-topic pairs to be judged should be carefully selected and the collected judgments should be consistent to achieve reliable evaluation. On top of these challenges, multi-task test collections impose additional challenges: both the topic set and collected judgments should be reusable across the different tasks, i.e., the topics represent real information needs in all tasks and the collected judgments should be reusable across the tasks to minimize the human effort and cost. Moreover, we aim to construct a collection that is large-scale, easily-accessible, and highly-available.

In this section, we outline our methodology to construct a multi-task tweet test collection that addresses those challenges and meets the target objectives. Section~\ref{buildcoll} provides more details on how we applied the methodology to construct \evetar{}, our Arabic test collection. 
\subsection{Significant Events as Topics}\label{sec_significant_events}
We elected to choose \emph{significant} (i.e., popular) events to be the core of our test collection. A significant event is defined as an occurrence that happens at a particular time in a specific location and is discussed by the media (e.g., covered by an online news article). Focusing on significant events has several advantages. First, it enables the development of topics that represent frequent information needs of Twitter users as they often search Twitter for timely and up-to-date information about events~\citep{teevan2011twittersearch, zhao2013questions}. Second, rich content (i.e., relevant tweets) is more available for popular events. Third, the popularity of the topics might  
help annotators to do more consistent judgments due to their probable familiarity with the significant events.
Finally, it inherently facilitates the support of multiple tasks that generally revolve around events, namely event detection, ad-hoc search, timeline generation, and real-time summarization, which are defined in more detail next. 

\subsection{Multi-Task Collection}
\label{sec:tasks}
With the choice of events to be at the heart of our test collection, 
\textbf{Event Detection (ED)} task is naturally supported. The task requires the detection of events over a \textit{stream} of tweets without prior knowledge on what events to expect and when they might happen. 
Our definition of an \emph{event} (given in Section~\ref{sec_significant_events}) 
is similar to definitions introduced in~\citep{mcminn2013building}; however, we emphasize event significance, which is often neglected in most event definitions~\citep{alsaedi2015arabic,becker_2012,petrovic2010streaming}. 
An event in our collection is \textit{represented} as a set of tweets that are relevant to it within a time period surrounding the time of that event. The event \textit{detected} by the event detection system would be represented as a subset (not necessarily all) of the tweets that are relevant to that event. 

ED task requires relevance judgments for each event. This inherently enables our test collection to support the \textbf{Ad-hoc Search (AS)} task as well.
Ad-hoc search is the typical search task in IR in which an ad-hoc query (representing a topic of interest) is issued at a search system which is required to retrieve a ranked list of documents (i.e., tweets) that are relevant to that topic over a collection of documents. The query is usually tagged with a time stamp that indicates the time at which it is issued. In \evetar{}, each topic is represented by multiple different queries that can be used in ad-hoc search, e.g., the title of the event.

As Twitter has become a valuable information source from which we can get instant updates about topics (or events) of interest, the massive amount of data posted on Twitter makes it very challenging to track the topics manually. Therefore, an IR system that helps users filter out redundant information and elect relevant updates in real-time is crucial.
The \textbf{Real-time Summarization (RTS)}
track was introduced in TREC 2016 to tackle this exact problem~\citep{lin2016overview}. 
According to TREC definition, an RTS system monitors a stream of tweets in real-time to detect relevant but non-redundant tweets for a given topic (called interest profile). Those tweets are then pushed (in real-time) to the mobile of the interested user. The pushed tweets represent the developed summary of the topic over time.
In TREC, the maximum number of tweets pushed per day is limited to 10 to avoid overwhelming users. Note that a relevant tweet can be redundant if another tweet covering the same information has already been pushed and RTS systems are indeed penalized when they return redundant tweets. Therefore, in addition to relevance judgments, we also need  \textit{novelty} judgments to support the RTS task. \evetar{} adopts the same exact definition but applied on events; however, the time for each event is limited to its active period only (5 days as explained in Section~\ref{buildcoll}), simulating the fact that the user would be interested in the event only during that period.

Having relevancy and novelty judgments, we can also support \textbf{Tweet Timeline Generation (TTG)} ~\citep{lin2014overview} task which is the retrospective version of the RTS task. Given a topic at a specific time, a TTG system is required to return a \textit{set} of relevant but non-redundant tweets in a chronological order (called a \textit{timeline}). The timeline is considered as a retrospective summary of the topic at the issuing time.

\begin{figure}[h]
\centering
\includegraphics[width=\textwidth]{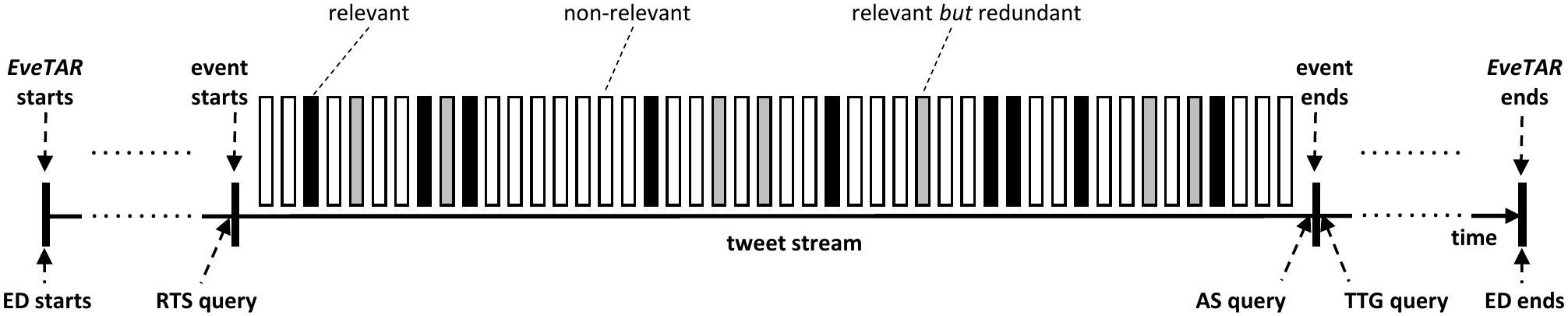}
\caption{The four supported retrieval tasks over \evetar{}. ED and AS tasks target relevant tweets (black and gray) while TTG and RTS tasks target relevant but not redundant tweets (black only)}
\label{fig:queries}
\end{figure} 

Both ED and RTS are considered streaming tasks since each processes the tweets as a stream; whereas AS and TTG are considered retrieval tasks since each processes the tweets as a collection. Figure~\ref{fig:queries} illustrates the relationship between the four tasks with respect to time and also type of targeted tweets for an example event. For that event, sets of relevant and also novel tweets are identified over its specified time period. 
The evaluation of both ED and AS tasks requires only the relevant tweets (black- and gray-colored), while the evaluation of both TTG and RTS requires relevant and novel tweets (black-colored).
The ED task starts with the beginning of the collection since all events should be detected, while RTS task per topic starts when the topic becomes of interest, i.e., when the event starts. Queries of both AS and TTG tasks are issued at the end of the event because they require retrospective retrieval of tweets.
Table~\ref{tasks} contrasts the four tasks with respect to the nature of the input and output, and whether the tasks consider the novelty of retrieved tweets. 

\begin{table}[h]
\caption{Supported IR Tasks\label{tasks}}
\begin{tabularx}{\textwidth}{XlcX}
\hline\noalign{\smallskip}
\textbf{Task}& \textbf{Input Data} & \textbf{Novelty} & \textbf{Expected Output}\\
\noalign{\smallskip}\Xhline{2\arrayrulewidth}\noalign{\smallskip}
    \textbf{Event Detection} & Tweet stream & \ding{55} & Events (represented by sets of tweets)\\
    \textbf{Ad-hoc Search} & Tweet collection & \ding{55} & Ranked list of tweets\\
    \textbf{Real-time Summarization} &  Tweet stream & \ding{51} & 10 tweets per day \\
    \textbf{Timeline Generation} & Tweet collection & \ding{51} & A set of tweets\\
\noalign{\smallskip}\hline
\end{tabularx}
\end{table}

\subsection{Large-Scale and Dense Dataset}
To properly support the evaluation of the four IR tasks, we need to acquire a tweet dataset that is both \emph{large-scale} and \emph{dense}. In other words, we need to ensure breadth, to cover a good number of significant events, and depth, to have enough relevant content per event. 

To achieve the needed breadth, the dataset has to span a time period that is long enough to cover several events. Considering that those events represent our topic set, we have to ensure the coverage of a number of events that is typically used in TREC to construct a \emph{reliable} test collection, which is roughly 50.
For Arabic collection, as an example, a month is expected to cover that number.

To achieve the needed depth, the Twitter's streaming API\footnote{\url{https://dev.twitter.com/streaming/reference/get/statuses/sample}} typically provides around 1\% random sample of tweets posted in Twitter and covering all languages used in the platform; however, tweets of a language that is not dominant (e.g., Arabic) will be an even lower percentage of that sample. To avoid sparseness, the API can then be used to track (and thus collect) tweets that include \textit{frequent} keywords, ensuring a focused set of tweets that are (almost) all of the target language, while abiding by Twitter's tweet crawling rate limit. Since the set of keywords being tracked controls which tweets will be added to the final dataset, a special attention has to be paid to the selection of those words in order to maximize the chance of obtaining a representative set of tweets of the target language on Twitter.
\subsection{High-Coverage and Diversified Judgment Pool}
In our test collection, we need two types of judgments, relevance and novelty; however, relevance judgments have to be collected first before determining which of the relevant ones are also novel. This requires getting first a set of potentially-relevant documents per topic that constitutes the judgment pool.

Ideally, each document-topic pair should be judged manually. However, due to the high cost of relevance judging,
we can judge only a subset of documents in practice. Clearly,
the choice of documents to be judged has a significant impact on the quality of the test collection. Traditionally, unjudged documents are assumed to be irrelevant; therefore, in collecting the relevance judgments for a topic, it is crucial to acquire relevant documents/tweets as much as possible. Furthermore, the method of selecting tweets to be judged should not be biased to a specific retrieval algorithm in order to achieve reliable evaluation. 
Many methods have been proposed to select documents to be judged, such as pooling~\citep{jones1975report}, statAP~\citep{pavlu2007practical}, MTC~\citep{carterette2006minimal}, and interactive judging~\citep{cormack1998efficient}. Most of these methods require a diverse set of systems to enrich the document pool. Developers of TREC test collections overcome this problem by running a shared-task evaluation campaign where research teams participate with their own systems.
On one hand, organizing a shared-task is a very costly process. On the other hand, developing many systems is also a very challenging process, especially for languages with limited IR resources such as Arabic.

\citet{moffat2015pooled} showed that query variations for a given topic are as strong as system variations in producing a diverse document pool.
Therefore, in order to diversify the tweet pool, we adopt a process that manually crafts a list of keyword and phrase queries for each topic
by performing \emph{interactive search} on Twitter's website. That (compared to searching our collection) would ensure a wider coverage of the topic aspects.
The crafted queries can then be used to search the collection using an off-the-shelf retrieval engine to retrieve the potentially-relevant documents that constitute the judgment pool.
\subsection{Reliable Judgments}
After identifying the potentially-relevant tweets, we need to collect relevance and novelty judgments. There are two issues to consider:
\begin{enumerate}
\item 
For relevancy, the pool for each topic can be annotated by several annotators, since the decision on whether a tweet is relevant is independent of the others. However, that is not the case for novelty judgments, where the decision on whether a tweet is novel depends on the other relevant tweets.
\item
Since the topics are significant events, we expect a relatively large pool of tweets to judge for relevancy, which calls for a large-scale annotation effort. However, the pool for novelty is expected be much smaller as it includes the relevant tweets only.
\end{enumerate}

Based on that, crowd-sourcing is a suitable option for relevancy annotations in terms of scale; however, we need to ensure the reliability and quality of the judgments. We achieve that through three steps: (1) we require multiple judgments per tweet, (2) We require a minimum quality level of annotators based on their work history, and (3) we employ pre-qualification test for potential crowd-sourcing annotators who are also required to maintain a minimum accuracy throughout the process. To further improve the overall quality of the annotations, after collecting the relevance judgments of all topics, we exclude the topics of low Kappa inter-annotator agreement~\citep{fleiss1971measuring}, keeping only the topics of high agreement.

The novelty annotations require the annotator to go through the entire stream of relevant tweets for a given topic to decide which of them are redundant and which are novel. We therefore opt to hire in-house annotators to ensure consistency and integrity of the annotations. This allows us to closely train the annotators and monitor their work.
\subsection{Easily-Accessible Highly-Available Collection}
Due to Twitter's tweet re-distribution policy, only tweet IDs (not content) can be distributed. Therefore, researchers interested in using the collection have to re-crawl the tweets from Twitter using their IDs. That introduces two major challenges. First, many tweets might become inaccessible over time, e.g., deleted tweets or tweets of deleted or suspended accounts. This will make performance comparisons over the collection less meaningful due to the use of probably different crawls performed at different times. 
Second, the estimated crawl time of a large set of millions of tweets over a single machine is in the order of tens of days based on the rate limit enforced by Twitter's API. 
This can be challenging for research teams that are interested in using the collection but do not have the time or computational power to crawl it.

To tackle the first challenge, we mitigate the effect of inaccessible tweets by removing (most of) them from the collection (i.e., not including them in the released version) after a reasonable period of time has been spent since the original crawl. That would make the collection more stable by reducing the chance of including deleted tweets for example, as we believe it decays over time. 

To tackle the second one, we make multiple smaller subsets available; all include the full judged set of tweets. The subsets have to be small enough to be easily crawled in a timely manner, while being representative of the complete collection. They also allow much faster evaluation. Moreover, they make it possible to conduct experiments on different scales of data, which might lead to different insights. 
\section{Constructing \sevetar}
\label{buildcoll}
\begin{figure}[h]
\centering
\includegraphics[width=0.95\textwidth]{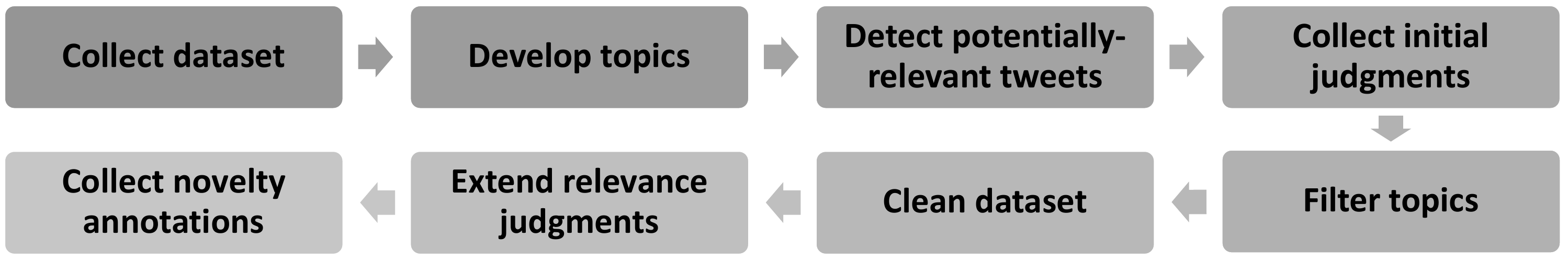}
\caption{Pipeline of steps followed to create \evetar{}}
\label{fig:pipe}
\end{figure}

Following the approach we presented in Section~\ref{approach}, we have constructed \evetar{}, a multi-task test collection over Arabic tweets. We have applied our approach over two phases. In the first phase, we collected a set of Arabic tweets, developed topics using Wikipedia's Current Events Portals, and collected relevance judgments via crowdsourcing. The collection at the end of this phase was released earlier~\citep{almerekhi2016evetar}. 
In the second phase, we aimed to improve the quality of the judgments, extend the collection to support more retrieval tasks, and make it more accessible.
We finally collected novelty labels to support TTG and RTS tasks. We also created four subsets of the collection to serve different purposes. In the following sub-sections, we describe in detail the pipeline of these steps illustrated in Figure~\ref{fig:pipe}.
\subsection{Collecting the Dataset}
To construct a test collection, a dataset has to be acquired first. We used the Twitter's streaming API\footnote{\url{https://dev.twitter.com/streaming/reference/get/statuses/sample}} to collect Arabic tweets spanning the period between December 30$^{th}$, 2014 and February 2$^{nd}$, 2015.
To avoid sparseness, we used Twitter's streaming API to \textit{track}\footnote{\url{https://dev.twitter.com/streaming/overview/request-parameters#track}} 400 frequent Arabic words so that we can collect Arabic tweets that match those words.
The words were identified as the most frequent in a set of 2M Arabic tweets that were crawled via the 1\% random sample of tweets collected over 10 days starting from April 10$^{th}$, 2014.
Using this methodology, we eventually collected a set of 590M Arabic tweets within the time period mentioned above.
\subsection{Developing Topics}
\label{ice}
The second major component of the test collection is the set of topics representing user needs. As mentioned earlier, we elected to select significant events to be the core topic set of \evetar{}. Guided by the work of~\citet{mcminn2013building}, we collected a set of 357 events that took place in January 2015 according to the English\footnote{\url{https://en.wikipedia.org/wiki/Portal:Current_events}} and Arabic\footnote{\url{http://bit.ly/2n5TYhY}} Wikipedia's Current Events Portals (WCEP). Each event in this list is represented by a date and a title (or an English-to-Arabic translated title). 
The event set went through two filtering phases to detect events meeting our significance criteria. In the first phase, we kept only the events that have been discussed by at least one online \emph{Arabic} news article, 
which resulted in a set of 71 events only. This process was done by manually searching for articles in popular Arabic news websites such as Aljazeera\footnote{\url{http://www.aljazeera.net}} and CNN Arabic\footnote{\url{arabic.cnn.com}} using event titles as queries. In the second phase, Twitter's advanced search service\footnote{\url{https://twitter.com/search-advanced}} was used to manually find at least 20 relevant Arabic tweets for each event using several manually-crafted Arabic queries per event. We limited the search for each event to a 5-day period starting 2 days prior to the event date (since some discussions about the event might start prior to it, while most of the relevant tweets can be found on the day of the event or within two days of its occurrence).
Following these two filtering steps, we acquired a final set of 66 significant events. Each event in \evetar{} is represented by a rich representation including event title, date, location, and category extracted from WCEP. Figure~\ref{extopic} shows a translated example of one event in \evetar{}. Other examples of events in \evetar{} include: ``North Korea hacks Sony accounts", ``Suicide bombing in Ibb", and ``Match of Australia vs Kuwait at the opening of the Asian Cup". 
\renewcommand{\arraystretch}{1.25}
\begin{figure}[h]
\centering
\small
  \begin{tabular}{|lp{5.9cm}|}
    \hline
     \textbf{ID}& E12\\
     \textbf{Title}& Discovery of tomb of Egyptian queen Khentakawess III\\
    \textbf{Date}& January 04, 2015\\
    \textbf{Location}&  Abusir, Egypt\\
    \textbf{Category}& Arts and Culture\\
    \textbf{Reference}& \url{http://cnn.it/1O6grQK}\\
    \textbf{Keywords}&Khentakawess, Egyptian queen, archaeologist\\
    \textbf{Description}&A Czech archaeological team discovered the tomb of an Egyptian queen named Khentakawess III who lived during the 5$^{th}$ dynasty. \\
    \hline
  \end{tabular}
  \caption{A translated example of an event as represented in \evetar{}\label{extopic}}
	\end{figure}
\subsection{Detecting Potentially-Relevant Tweets}\label{section_detecting_tweets} 
Having acquired the dataset and developed the topics, the next natural step is to collect judgments per topic. In \evetar{}, we have two types of judgments, relevance and novelty; however, relevance judgments have to be collected first before determining which of the relevant ones are also novel. 
To construct the judgment pool, we adopted an interactive search approach by manually crafting a list of keyword and phrase queries for each topic using the search service on Twitter's Website. 
For example, some of the (translated) queries for the event described in Figure~\ref{extopic} are: ``Abusir", ``Khentakawess Third" and ``Egyptian queen". 
Given an average of 6 queries per topic, we created one long ``OR'' query comprising all queries per topic, and retrieved 10K tweets per topic using Lucene open-source search library\footnote{\url{https://lucene.apache.org}}. We used Lucene's default retrieval model for ranking. We also limited the search to the event specific time period. After removing exact tweet-text duplicates, we obtained a pool of 134K tweets to be judged.
\subsection{Collecting Initial Relevance Judgments}
To collect relevance judgments for the pool, we used CrowdFlower crowdsourcing platform\footnote{\url{http://www.crowdflower.com}}. We conducted several pilot studies to design our tasks. We ran a separate judging task for each event/topic.
In each task, the corresponding title, description, and date of the event have been shown to annotators. We also provided the content of an Arabic news article discussing the event to help annotators better understand the topic and make better-informed judgments. We decided to display the content of the article rather than just providing a hyperlink to it, to increase the chance of annotators reading the article. 

Annotators were required to be Arabic speakers with an intermediate level based on CrowdFlower's ranking of annotators\footnote{\url{https://success.crowdflower.com/hc/en-us/articles/115000832063-Frequently-Asked-Questions}}. 
Before starting the judging process, we conducted a qualification test in which annotators were required to correctly label 8 out of 10 gold tweets. Gold tweets are randomly sampled from the set of candidate tweets per event and labeled by one of the authors.
Annotators were also required to maintain a minimum accuracy of 80\% over gold tweets embedded within the actual task. Eight annotators on average per task have failed to maintain that judging accuracy and were eliminated (along with all of their judgments). 

Each tweet-event pair was judged by 3 annotators to ensure a majority label. Annotators spent 59 judging hours per event on average. Given the 3 judgments per tweet, we used majority voting to select the final label. Eventually, our relevance judgment set consisted of 134K labeled tweets, out of which 51K tweets were judged as relevant. 
We measured inter-annotator agreement per event using Fleiss' Kappa~\citep{fleiss1971measuring}, which allows measuring agreement when a data item is annotated by more than 2 annotators. The average Kappa value over all topics is 0.60. The dataset, 66 topics, and collected judgments (at that point) were made available in an early version of \evetar{} 
~\citep{almerekhi2016evetar}.
\subsection{Filtering Topics}
To improve the average judgment quality of \evetar{}, we only kept the 50 topics with the highest agreement levels, dropping the lowest 16.
The distribution of categories of the 50 events is shown in Table~\ref{tab:dist:2}. As the table shows, events in the ``Armed conflicts and attacks" category constitutes 64\% of the entire set; this is due to the distribution of events in WCEP itself. 

\begin{table}[h]
\caption{Distribution of events in \evetar\label{tab:dist:2}}
  \begin{tabularx}{\textwidth}{XX}
\hline\noalign{\smallskip}
 \textbf{Category (Events)} & \textbf{Category (Events)}\\
\noalign{\smallskip}\Xhline{2\arrayrulewidth}\noalign{\smallskip}
   Armed Conflicts \& Attacks (32) &  Sports (5)\\
     Business \& Economy (1) & Arts \& Culture (2) \\
  International Relations (3) & Law \& Crime (1) \\
 Disasters \& Accidents (2) & Politics \& Elections (4)\\
\noalign{\smallskip}\hline
  \end{tabularx}
	\end{table}

\begin{figure}[h]
\centering
\includegraphics[width=\textwidth]{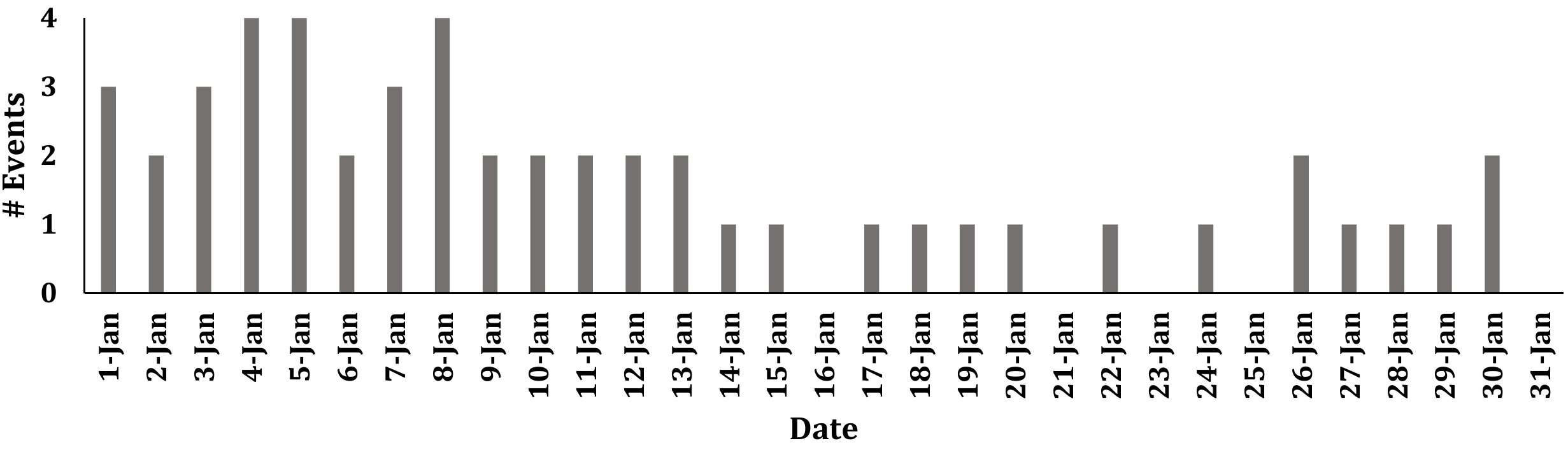}
\caption{Daily distribution of \evetar{} topics over the month of January, 2015}
\label{fig:tempdist}
\end{figure}

We also investigate whether the 50 events are temporally biased. Figure~\ref{fig:tempdist} shows the number of \evetar{} events started on each day of January 2015. We can see that the events  
are generally well-distributed over the period of the collection indicating that \evetar{} events are not biased to specific dates.

\subsection{Cleaning the Dataset}\label{section_cleaning}
Tweets might become inaccessible over time due to various reasons~\citep{bagdouri_2015}. Therefore, removing inaccessible tweets from the test collection is essential to evaluate IR systems reliably over a re-crawled version of \evetar{}, and also reduce re-crawling time. 

In December 2016, we examined the judged tweets released with the early version of \evetar{}~\citep{almerekhi2016evetar} and found that around 40\% of them have been inaccessible. The two major reasons behind that were (1) due to users deleting their own tweets, or (2) the whole user account is no longer accessible because either it was suspended\footnote{Shut-down by Twitter; check \url{https://support.twitter.com/articles/15790}.}, deleted, or made private by the user. The majority resulted from the inaccessibility of accounts (which is inline with  findings of Liu et al.'s study~\citep{ICWSM148043} conducted at a very large scale).

Due to the huge size of the dataset (590M tweets), it was infeasible to automatically check each tweet for deletion. 
As a compromise, we opted to apply an efficient solution; we observed that 
 the majority of the inaccessible tweets belong to inaccessible accounts. Therefore, we have automatically checked\footnote{on December 25$^{th}$ 2016.} each of 7.1M unique accounts in our original dataset and only kept the accessible accounts along with their tweets, resulting in around 5.8M accounts and a final dataset of 355M tweets.
\subsection{Extending Relevance Judgments}
Cleaning the dataset has encouraged us to improve the relevance judgments as well.
We first removed retweets (since they express no new information) 
and filtered out all inaccessible tweets. That resulted in a significant drop in the number of relevance judgments.
We also observed that many of the labeled tweets were written in modern standard Arabic (MSA); however, Twitter Arab users tend to use dialectal Arabic in their tweets~\citep{darwish2014arabic}. Therefore, we decided to extend the relevance judgments and increase linguistic diversity among labeled tweets. 

To achieve that, we developed 3 to 9 new queries per topic following the same earlier guidelines while attempting to include dialectal queries whenever possible. 
In a similar approach to that explained in Section~\ref{section_detecting_tweets}, 
we used the developed queries in searching using Lucene over an index of the cleaned dataset.
For each topic, the top 1K tweets were considered for labeling. Before passing those tweets to the relevance labeling task, we applied exact-text deduplication and retweet removal, and also excluded the tweets that have been judged before. 
Eventually, we collected 180 additional judgments on average per topic using the same crowdsourcing tasks we employed previously. 

Since some of the labeled tweets have exact duplicates in the collection, a system that returns an exact duplicate rather than the labeled tweet would not be scored on that tweet. Moreover, a recent study~\citep{Baruah_2014} has found that adding exact duplicates to the judgment set significantly affects system performance over different queries for 13 out of 28 systems participated in the temporal summarization track in TREC 2013~\citep{ts2013overview}. Therefore, we also propagated the relevance judgments by adding 3,947 exact duplicates of judged tweets fetched from the dataset. 
Note that we only considered exact duplicates that appear within active period of each event. 

\begin{figure}[h]
\centering
\includegraphics[width=0.9\textwidth]{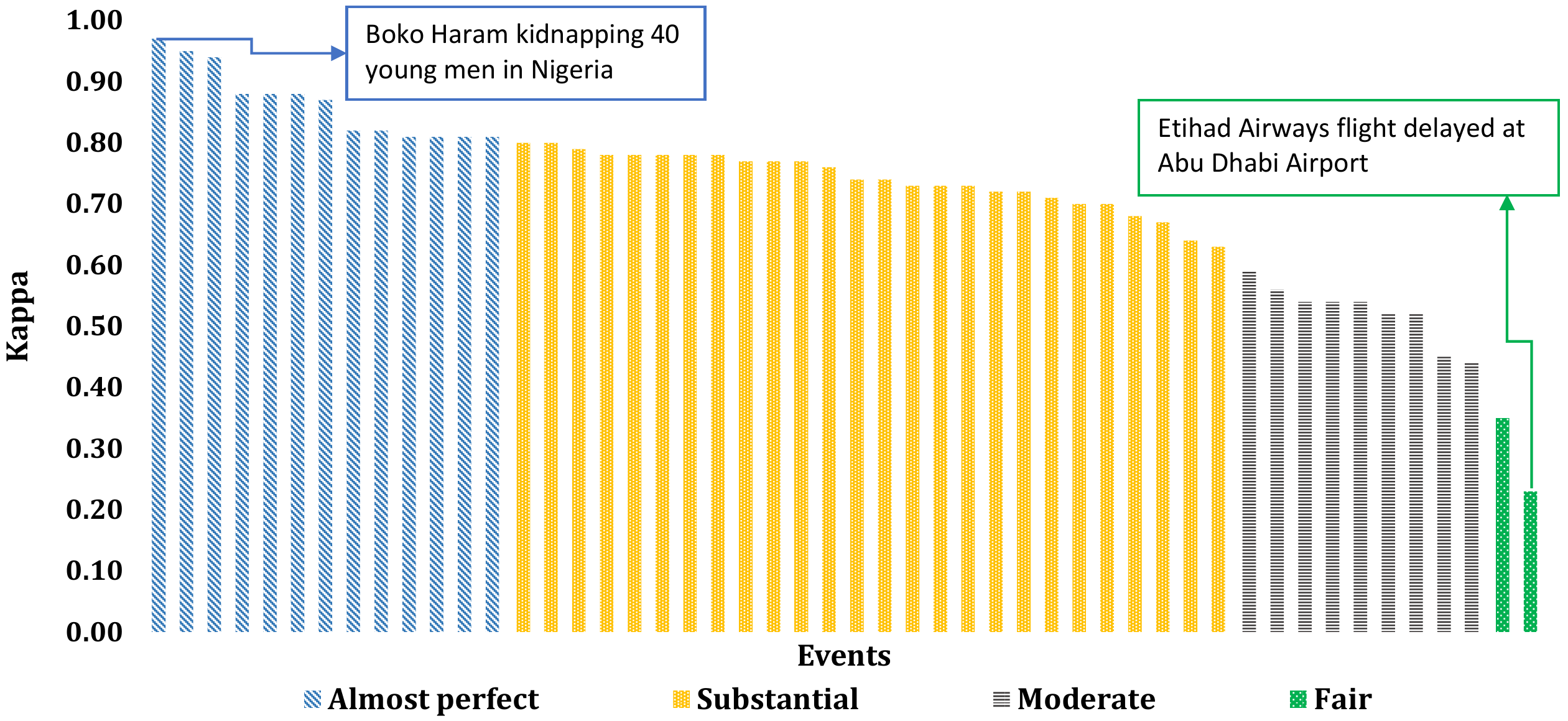}
\caption{Distribution of \evetar{} events over categories of Kappa values}
\label{fig:kcat}
\end{figure}

Eventually, we have a final relevance judgment set of 61,946 tweets, out of which 24,086 (39\%) are relevant. Computing Fleiss' Kappa over the final qrels resulted in an average Kappa of 0.71, which is considered a \emph{substantial} agreement according to a widely-adopted Kappa categorization~\citep{sim2005kappa}, showing how strong the agreement is. Figure~\ref{fig:kcat} illustrates the distribution of the 50 events according to that categorization. 78\% of the events has an inter-annotator agreement that is above moderate which indicates a strong inter-annotator agreement for most of the topics in \evetar{}. We also observe that 2 topics only got an agreement as low as fair. Analyzing labels for the topic with the least agreement, we observe that conflicts in labels might have happened because of incompleteness in the title of the event. Although the topic is about a delay of a specific Etihad Airways flight to San Fransisco, the title did not reflect the important piece of information about the destination of the flight. At the same time, the article displayed to the annotators included that piece of information. For further studies on crowdsourcing, we released the complete set of judgments and inter-annotator agreement per event.

\begin{figure}[h]
\centering
\includegraphics[width=\textwidth]{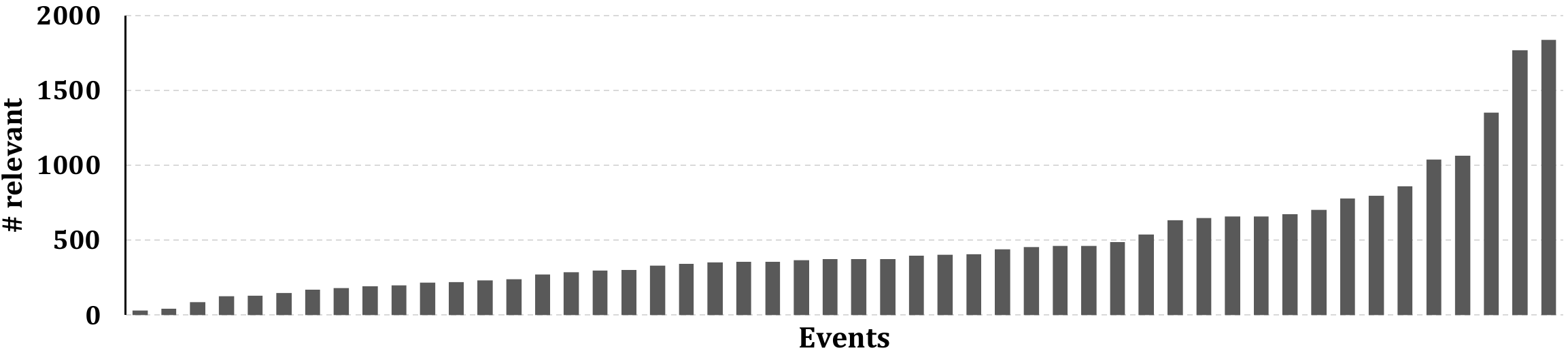}
\caption{Distribution of relevant tweets over \evetar{} topics}
\label{fig:reldist}
\end{figure}

Figure~\ref{fig:reldist} shows the number of relevant tweets per event. The figure demonstrates that the size of the relevant sets spans a large range and that we have events at all difficulty levels, 
assuming that events with more relevant judgments are easier to handle. 

\subsection{Collecting Novelty Annotations}
\label{novelty}


Novelty judgments are considered a second layer of labels on top of the relevance judgments. To collect the novelty judgments, we adopted the following definition of semantic-similarity between on-topic tweets, introduced in TREC 2015 Microblog~\citep{lin2015overview} and TREC 2016 RTS~\citep{lin2016overview} tracks. Relevant tweets for each topic are distributed into clusters of ``semantically-similar'' tweets; each cluster has a set of relevant tweets considered to carry the same information. Particularly, while receiving a chronologically-ordered stream of relevant tweets for a given topic, any incoming tweet that conveys the same information as previously-seen tweets, or it contains additional information that the user will not be interested to see, is considered redundant and is added to the cluster of previously-seen similar tweets. Otherwise, the tweet is judged as novel and it initiates a new cluster.

Following that definition, we collected novelty judgments using a web-based annotation interface~\citep{wang2015assessor}. 
We recruited 12 in-house annotators (a mix of undergraduate students and alumni of Qatar University)
and conducted two 3-hour training sessions, each covered motivation, task definition, and a demo of the annotation interface. Each annotator had to annotate two training topics before being eligible to work on the task. After checking the quality of annotations on the training topics, only 9 annotators were finally selected to work on the task.
We made the instructions of the task available after the training session
\footnote{\url{https://reemsuwaileh.github.io/EveTARNovelty/training.html}} for annotators to refer to during the actual task. 

After excluding the exact duplicates of relevant tweets, only 22K relevant tweets had to be annotated for novelty over 50 topics\footnote{Cluster labels were eventually propagated back to duplicate tweets.}. 
Each topic was annotated by one annotator. The number of topics each annotator worked on varied from 1 to 9 depending on annotators' speed and availability, and also topic pool size. The whole annotation task finished in approximately 100 hours over 3 weeks and produced 66 clusters per topic on average.  



\subsection{Subsets}\label{setup}
The estimated crawl time of 355M tweets over a single machine is 41 days based on the rate limit enforced by Twitter's API on tweets re-crawl\footnote{\url{https://dev.twitter.com/rest/reference/get/statuses/lookup}}. 
Therefore, it can be challenging to re-crawl \evetar{} by research groups with limited computational resources.
In order to make \evetar{} more accessible, 
we created a 
subset that can be re-crawled in a shorter 
time period 
while supporting the evaluation on different scales of datasets. 
Furthermore, the Arabic language has many dialects that are also commonly used on Twitter~\citep{darwish2014arabic}. To support the evaluation and performance comparisons of IR systems over dialectal vs. MSA tweets, we also provided the MSA and dialectal subsets of the sample. Finally, we provided another subset that includes only qrels to support even faster  
evaluation and facilitate quality analysis of crowdsourcing annotations. 
\begin{table}[h]
\caption{EveTAR Test Collection Versions\label{versions}}
\begin{tabularx}{\textwidth}{XXrrr}
\hline\noalign{\smallskip}
    \multicolumn{1}{X}{\textbf{Version}} & \multicolumn{1}{X}{\textbf{Description}} & \multicolumn{1}{c}{\textbf{Size}} & \multicolumn{1}{c}{\textbf{\#qrels}} & \multicolumn{1}{c}{\textbf{\#rels}}\\
\noalign{\smallskip}\Xhline{2\arrayrulewidth}\noalign{\smallskip}
    \textbf{\evetar{}-F} & Full collection & 355,821,033 & 61,946 & 24,086\\
    \textbf{\evetar{}-S} & Random sample & 15,057,208 & 61,946 & 24,086\\
    \textbf{\evetar{}-S.m} & MSA tweets of the sample & 8,066,583 & 47,369 & 21,233\\
    \textbf{\evetar{}-S.d} & Dialectal tweets of the sample &  6,990,533 & 14,577 & 2,853\\
    \textbf{\evetar{}-Q} & Qrels only& 59,732 & 61,946 & 24,086\\
\noalign{\smallskip}\hline
\end{tabularx}
\end{table}

Table~\ref{versions} summarizes the different versions of \evetar{} we made publicly available. In addition to \evetar{}-F, which is the full collection of 355M tweets, we also release the following four subsets: 
\begin{enumerate}
\item \textbf{\evetar{}-S}: A random sample of 15M tweets of \evetar{} (i.e.,\ 4\% of the full dataset) that includes all judged tweets and can be recrawled in 1.5 days based on our estimation. 
\item \textbf{\evetar{}-S.m}: The subset of \evetar{}-S that includes only MSA tweets.
\item \textbf{\evetar{}-S.d}: Similar to \evetar{}-S.m but  includes only dialectal tweets.
\item \textbf{\evetar{}-Q}: The subset of the full collection that  includes only the judged tweets (i.e., qrels). Notice that the size of that subset is smaller than the size of qrels because one tweet might be judged for more than one event.
\end{enumerate}

We classified tweets of \evetar{}-S into MSA and dialectal subsets using a pre-trained classifier~\citep{eldesouki2016qcri} which achieved the best accuracy (83.9\%) in DSL shared-task\footnote{\url{http://ttg.uni-saarland.de/vardial2016/dsl2016.html}}. The classifier predicted 46.4\% of the random sample as dialectal tweets. 
Among tweets of \evetar{}-S, there were 92 tweets that contain no text, and thus were filtered out and not included in neither of the two subsets.
\section{Evaluation}
\label{eval} 
In this section, we evaluate \evetar{} from three perspectives. First, we demonstrate its usability 
for evaluating the tasks it supports (Section~\ref{section_using_evetar}). This also produces reference performance results for future advancements in those tasks. 
Second, we evaluate the reliability of the test collection in ranking IR systems in  different tasks  
and investigate the effect of sampling in terms of evaluation reliability  
(Section~\ref{section_reliability}). Finally, we compare \evetar{} with other widely-used TREC microblog test collections (Section~\ref{section_putting_in_context}). 
\subsection{Usability}\label{section_using_evetar}

We first evaluate strong existing IR systems for each supported task using \evetar{}. Experiments are conducted over each version of \evetar{} (whenever feasible) to provide baseline performance results at different scales.
\subsubsection{Event Detection}
As events are at the heart of \evetar{}, event detection is a primary supported task.  
We ran two off-the-shelf event detection algorithms on \evetar{} to demonstrate its usability for this task. We used open-source implementation of Trending Score~\citep{benhardus2013streaming} and Peaky Topics~\citep{shamma2011peaks} algorithms provided by SONDY platform\footnote{\url{mediamining.univ-lyon2.fr/people/guille/sondy.php}}. 
We used SONDY's default parameter settings and ran both over all versions of \evetar{} except the full dataset as we did not have enough computational power to apply the algorithms at that scale. 

To automatically evaluate the algorithms, we adopted  Petrovic's~\citep{petrovic2013real} approach which measures effectiveness using recall:
\begin{equation}
recall = \frac{\#\text{\textit{covered events}}}{\#\text{\textit{reference events}}}
\end{equation}
where $\#\text{\textit{reference events}}$ is the number of events in \evetar{} and $\#\text{\textit{covered events}}$ is the number of reference events covered by the algorithm's detected events (where an event is represented by a list of tweets). 
A reference event is assumed to be covered by a detected event if at least 50\% of the tweets in the detected event belong to that reference event.
It should be noted that it is not feasible to automatically compute precision under this setup, since the ``complete'' list of possible significant events in the collection is not available. 
As it is difficult to manually compute precision given the huge number of events detected by each algorithm, we only report recall results in Table~\ref{tab:Table5}. 
\begin{table}[h]
\caption{Event detection over \evetar{}. Best recall per version is \textbf{boldfaced}\label{tab:Table5}}
\begin{tabularx}{\textwidth}{XYY}
\hline\noalign{\smallskip}
{\bf Version} & {\bf Peaky Topics} & {\bf Trending Score} \\
\noalign{\smallskip}\Xhline{2\arrayrulewidth}\noalign{\smallskip}
\evetar{}-S & 0.42 & \textbf{0.52} \\
\evetar{}-S.m & 0.74 &\textbf{ 0.92}  \\
\evetar{}-S.d & 0.14 & \textbf{0.48} \\
\evetar{}-Q & \textbf{1.00} & \textbf{1.00} \\
\noalign{\smallskip}\hline
\end{tabularx}
\end{table} 

The table shows that Trending Score is outperforming Peaky Topics over all versions of \evetar{}. We also observe that both algorithms managed to achieve perfect recall over \evetar{}-Q due to its small size and the very high prevalence of relevant documents. Similarly, event detection over \evetar{}-S.m is noticeably better than that over \evetar{}-S and \evetar{}-S.d due to the higher prevalence of relevant documents in \evetar{}-S.m (0.26\%) as opposed to the other two subsets (0.16\% in \evetar{}-S and 0.041\% in \evetar{}-S.d).
\subsubsection{Ad-hoc Search}\label{adhoc_systems}
Using \evetar{}, we evaluated four AS systems, denoted by query likelihood (QL), query expansion (QE), temporal decay (TD) and temporal relevance modeling (TRM), which were adopted by one of the top teams~\citep{hasanain2014qu} in the ad-hoc search task in TREC 2014 microblog track~\citep{lin2014overview}. Moreover, these models represent a diverse set of retrieval models usually used in tweet search.
To be able to run those systems on Arabic tweets, we added Arabic-specific preprocessing modules, e.g., stemming and removing stopwords, emoticons, and diacritics, in addition to URL and mentions removal. We ran the systems (using their reported parameter values) on all 5 versions of \evetar{}, as shown in Table~\ref{tabadhoc}, and limited the search to the active period of each event.
We used mean average precision (MAP) and precision at rank 30 (P@30) averaged over all 50 topics to evaluate system performance, as typically used~\citep{lin2014overview}. 

\begin{table}[h]
\caption{Ad-hoc search over \evetar{}. Best result per version per evaluation measure is \textbf{boldfaced}\label{tabadhoc}}
\begin{tabularx}{\textwidth}{lYYYY@{\extracolsep{12pt}}YYYY}
\hline\noalign{\smallskip}
& \multicolumn{4}{c}{\textit{MAP}}& \multicolumn{4}{c}{\textit{P@30}} \\
\cline{2-5}\cline{6-9}\noalign{\smallskip}
\bf Version & \bf QL & \bf QE & \bf TD & \bf TRM & \bf QL & \bf QE & \bf TD & \bf TRM \\
\noalign{\smallskip}\Xhline{2\arrayrulewidth}\noalign{\smallskip}
\evetar-F & 0.3951 & 0.4494 & 0.3926 & \textbf{0.4534} & 0.7340  & 0.7180 & 0.7313  & \textbf{ 0.7707} \\
\evetar-S & 0.5986 & \textbf{0.6642} & 0.5961 & 0.6630 & 0.8373 & 0.8520 & 0.8360 & \textbf{0.8747} \\
\evetar-S.m & 0.6202 & 0.6848 & 0.6184 & \textbf{0.6946} & 0.8473 & 0.8480 & 0.8453 & \textbf{0.8680} \\
\evetar-S.d & 0.5529  & 0.5817 & 0.5505 & \textbf{0.6006} & 0.5007 & \textbf{0.5313} & 0.4927 & 0.5233 \\
\evetar-Q & 0.6340 & 0.6882 & 0.6317 & \textbf{0.6951} & 0.8133 & \textbf{0.8600} & 0.8140 & 0.8373 \\
\noalign{\smallskip}\hline
\end{tabularx}
\end{table}

Table~\ref{tabadhoc} shows that, in both measures, TRM is generally outperforming all other models in all versions of \evetar{}. As expected, QE is almost as effective as TRM since both systems are query expansion retrieval models that utilize pseudo-relevant tweets to expand the query.
Results also showed that the systems are not as effective over \evetar-F and \evetar-S.d as they are over the other subsets. This is due to the low prevalence of relevant documents in these two subsets (0.007\% in the full version and 0.041\% in the dialectal sample) as opposed to the other subsets, making it a more difficult retrieval task.
\subsubsection{Tweet Timeline Generation} \label{section_ttg}
We conducted experiments with two TTG systems, cutoff and incremental clustering, adopted by one of the top teams~\citep{hasanain2014qu} in the TTG task in TREC 2014 microblog track~\citep{lin2014overview}. Both systems follow a two-step approach to generate a timeline. The system first retrieves a set of candidate tweets in response to a topic using a retrieval model, then applies a summarization technique to the retrieved set to generate a timeline. In the first step, we experimented with the same four ad-hoc retrieval models presented earlier, namely QL, QE, TD, and TRM. 

We enabled the TTG systems to run on Arabic tweets similar to the AS systems, and configured both using their reported parameter values. We used weighted $F_1$ ($wF_1$) measure, which is the official one used in TREC 2014~\citep{lin2014overview}, to evaluate the performance of the systems. 
\begin{table}[h]
\centering
\caption{TTG over \evetar{}\label{tabttg}. The best result per version is \textbf{boldfaced}}
\begin{tabularx}{\textwidth}{lYYYY@{\extracolsep{12pt}}YYYY}
\hline\noalign{\smallskip}
& \multicolumn{4}{c}{{\bf Cutoff}} & \multicolumn{4}{c}{{\bf Incremental Clustering}} \\
\cline{2-5}\cline{6-9}\noalign{\smallskip}
\bf Version&  \bf QL & \bf QE & \bf TD &\bf TRM &\bf QL &\bf QE &\bf TD &\bf TRM \\
\noalign{\smallskip}\Xhline{2\arrayrulewidth}\noalign{\smallskip}
\evetar-F &     0.2164 &     0.2009 &     0.2180 &     0.2043 &     0.3649 &     0.3578 &     0.3608 &     \textbf{0.3809} \\
\evetar-S &     0.2574 &     0.2474 &     0.2588 &     0.2450 &     0.4301 &     0.4210 &     0.4296 &     \textbf{0.4329} \\
\evetar-S.m &     0.2586 &     0.2474 &     0.2597 &     0.2563 &     0.4252 &     0.4144 &     0.4265 & \textbf{0.4374} \\
\evetar-S.d &     0.1817 &     0.1874 &     0.1793 &     0.1879 &     0.2229 & \textbf{0.2337} &     0.2203 &     0.2325 \\
\evetar-Q &     0.2648 &     0.2445 &     0.2686 &     0.2520 &     0.4070 &     0.3819 &     0.4058 &     \textbf{0.4115} \\
\noalign{\smallskip}\hline
\end{tabularx}
\end{table}

Table~\ref{tabttg} reports the average $wF_1$ scores over the 50 topics for both TTG systems over \evetar. Similar to AS results, the table shows that TTG systems are less effective with \evetar-F and \evetar-S.d test collections, due to lower percentage of relevant tweets. We also notice that  Incremental Clustering is far more effective than Cutoff. Combining Incremental Clustering with the TRM retrieval model achieves the best performance over almost all versions of \evetar{}. 
This is consistent with findings of a recent study that showed the performance of TTG systems adopting the 2-step approach is highly dependent on the underlying retrieval model~\citep{Magdy2016}. 


\subsubsection{Real-Time Summarization}\label{section_rts}
We evaluated three RTS systems over \evetar{}. Two of them, namely Vector Space Model (\textit{VSM}) and \textit{Rocchio}~\citep{suwailehqu2016}, constitute the best two automatic systems in the ``push notification'' scenario in TREC 2016 RTS track (denoted then as QUBaseline and QUExpP). We also adopted a third system, denoted as \emph{Silent}, that was designated as the baseline at the same track~\citep{lin2016overview}. 
Basically, this system is always silent, i.e., not pushing anything to the user during the entire evaluation period and just accumulates the gain of silent days in which a topic does not have any relevant tweets.

\begin{table}[h]
\centering
\caption{RTS over \evetar\label{tab:rts}. The best result per version per evaluation measure is \textbf{boldfaced}}
\begin{tabularx}{\textwidth}{lYYY@{\extracolsep{12pt}}YYY@{\extracolsep{12pt}}YY}
\hline\noalign{\smallskip}
 & \multicolumn{3}{c}{\bf \textit{EG-1}} & \multicolumn{3}{c}{\bf \textit{nCG-1} } & \multicolumn{2}{c}{\bf \#pushed} \\
\cline{2-4}\cline{5-7}\cline{8-9}\noalign{\smallskip}
  \bf Version & \bf VSM & \bf Rocchio & \bf Silent & \bf VSM & \bf Rocchio & \bf Silent & \bf VSM & \bf Rocchio\\
\noalign{\smallskip}\Xhline{2\arrayrulewidth}\noalign{\smallskip}
   \evetar{}-F & \textbf{0.2469} & 0.2333 & 0.1600 & \textbf{0.2688} & 0.2384 & 0.1600 &  466 & 316 \\
    \evetar{}-S & \textbf{0.2557} & 0.2365 & 0.1600 & \textbf{0.2799} & 0.2455 & 0.1600 &  392 & 251\\
    \evetar{}-S.m & \textbf{0.2766} & 0.2620 & 0.1800 & \textbf{0.2975} & 0.2620 & 0.1800 & 344 & 227 \\
   \evetar{}-S.d & \textbf{0.4737} & \textbf{0.4737} & 0.4320 & \textbf{0.4813} & 0.4741 & 0.4320 & 128 & 100\\
    \evetar{}-Q & \textbf{0.2569} & 0.2362  & 0.1600 & \textbf{0.2807} & 0.2479 & 0.1600 & 379 & 245 \\
\noalign{\smallskip}\hline
  \end{tabularx}
	\end{table} 
Table~\ref{tab:rts} illustrates the performance of the three systems measured by Expected Gain (EG-1) and Normalized Cumulative Gain (nCG-1)~\citep{lin2016overview} over all versions of \evetar{}. It also shows the total number of pushed tweets in each case. We can see that VSM outperforms other systems in all settings. In addition, EG-1 and nCG-1 scores of all systems in all versions of \evetar{} are relatively close to each other, except for \evetar{}-S.d. Upon investigation, we realized that the titles of topics used in summarization, were written in MSA, while tweets of that version are all in dialectal Arabic. Therefore, fewer relevant tweets have been detected as relevant and pushed by both VSM and Rocchio.
Since RTS evaluation is precision-oriented, systems pushing fewer but relevant tweets are more likely to achieve higher performance. 
Silent also achieves higher performance on \evetar{}-S.d, as there are more silent days in \evetar{}-S.d due to fewer relevant tweets in that subset. 
\subsection{Reliability}\label{section_reliability}


In the previous section, we demonstrated the usability of \evetar{} by evaluating the performance of multiple systems over the different supported tasks. In this section, we adopt Generalizability Theory (GT)~\citep{brennan2001statistics,bodoff2007test,Bodoff20081117,urbano2013measurement} to measure the reliability of the reported results.
GT is based on the analysis of variance components (i.e., variance in effectiveness scores due to systems, topics and system-topic interactions). 
We computed the \emph{generalizability coefficient} (GC) (using the implementation of~\citet{urbano2013measurement}) which measures the reliability of ranking of IR systems. For a reliable ranking, GC score approaches 1. Note that the GC measure is sensitive to systems used in the computation, and thus is used to measure the reliability of a specific ranking, not the overall quality of a test collection. 

In computing GC, we used MAP for the AS task, $wF_1$ for the TTG task, and EG-1 for the RTS task\footnote{Evaluation reliability of ED task is not reported because we need evaluation scores for each topic in order to apply GT. However, in evaluation of ED systems, we calculate a single score for all topics.}. We considered both \evetar{}-F and \evetar{}-S to investigate the effect of sampling in terms of evaluation reliability. To have a reference point, we computed GC scores of the rankings of same systems but over the test collections of the most recent TREC track of each task (TREC 2014 Microblog Track for AS and TTG, and TREC 2016 RTS Track for RTS).
Results presented in Table~\ref{table_reliability} show that \evetar{} achieves higher GC scores than TREC collections, which indicates that the reliability of \evetar{} in ranking the  used systems is at least comparable to TREC test collections. 

\begin{table}[h]
\centering
\caption{Generalizability Coefficient Scores of system rankings
\label{table_reliability}}
\centering
  \begin{tabularx}{\textwidth}{XYYY}
\hline\noalign{\smallskip}
    \textbf{Task} 	& \textbf{\evetar{}-F} & \textbf{\evetar{}-S}	& \textbf{TREC}  \\ \noalign{\smallskip}\Xhline{2\arrayrulewidth}\noalign{\smallskip}
    \textbf{AS}  & 0.937 	& 0.966	& 0.901  	\\ 
    \textbf{TTG} 	& 0.974 & 0.979 & 0.872    \\
    \textbf{RTS} 	& 0.974 & 0.975   & 0.743  \\
    \noalign{\smallskip}\hline
  \end{tabularx}
	\end{table} 

Table~\ref{table_reliability} also shows very similar GC scores for \evetar{}-S and \evetar{}-F, which suggests that sample subset is as reliable as the full collection in ranking systems. To further investigate the representativeness quality of the sample, we computed Pearson's correlation between performance scores of each system per topic over both \evetar{}-F and \evetar{}-S, and then averaged it over all systems per task\footnote{We are not able to report results for event detection since we could not run event detection on the full dataset due to its large size.}. Table~\ref{tab:corr} shows very high correlation between scores over both collections with very small standard deviation. Moreover, correlations for all systems are statistically significant at a 95\% confidence level. This suggests that, overall, the sampled version of \evetar{} (i.e.\ \evetar{}-S) is as reliable as the full version in evaluating the tasks when it is not feasible to acquire the full collection.

\begin{table}[h]
	\caption{Pearson's correlation of performance over \evetar-F and \evetar-S
\label{tab:corr}}
  \begin{tabularx}{\textwidth}{Xc}
\hline\noalign{\smallskip}
{\bf Task} &  \textbf{Average Pearson's Correlation (Standard Deviation)} \\ 
\noalign{\smallskip}\Xhline{2\arrayrulewidth}\noalign{\smallskip}
{\bf AS} & 0.866 (0.009)   	\\
{\bf TTG} &   0.892 (0.033)  	\\
{\bf RTS} & 0.973 (0.026) 		\\ 
\noalign{\smallskip}\hline
\end{tabularx}
\end{table}

\subsection{In Context of TREC Collections} \label{section_putting_in_context}


TREC test collections are among the most widely used test collections, covering a large variety of IR tasks. In this section, we compare \evetar{} (mainly the full version) against TREC microblog test collections used in TREC 2011-2015 Microblog tracks~\citep{ounis2011overview,soboroff2012overview,lin2013overview,lin2014overview,lin2015overview} and TREC 2016 RTS track~\citep{lin2016overview}. Table~\ref{tab:comp:trec:topics} summarizes the comparison in several aspects. 
\begin{itemize}
\item 
\textbf{Language:} \evetar{} has the largest monolingual tweet collection compared to TREC collections which target a smaller subset of English tweets to match the language of topics.
 Moreover, \evetar{} has MSA and dialectal subsets that can be useful in IR studies investigating linguistic diversity within the same language (e.g., cross-dialect IR). 
\item 
\textbf{Tasks:} \evetar{} supports four different IR tasks, one of them (ED) has never been supported by any of the TREC tweet test collections. 
\item 
\textbf{Topics:} The topic set size of \evetar{} is comparable to TREC collections. \evetar{} topics are defined in the TREC style (title, description, and narrative) in addition to having multiple queries per topic. Those queries can be useful in several research studies, e.g., automatic evaluation (e.g., \citep{efron2010query}) and development of new evaluation measures (e.g., \citep{zuccon2016query}).
\item 
\textbf{Qrels:} To create a pool of tweets to be judged, we used interactive search to generate multiple queries representing the same topic, which were then employed to search the collection and retrieve potentially-relevant tweets. TREC test collections employed the traditional pooling method via the shared-task campaign. While the pool size in \evetar{} is comparable to most of the TREC collections, the total number of relevant tweets (and also per topic) in \evetar{} is much higher than others. We expect this is due to two factors. First, the topics represent ``significant events'', which are naturally popular and more widely discussed. Second, \evetar{} is a much larger collection than the others. However, percentage of documents in the collection that are relevant is within the range observed in TREC collections. Finally, the high percentage of relevant tweets in the pool (39\%) might indicate the effectiveness of the multiple-queries in capturing relevant tweets. 
\end{itemize}
The above comparison suggests that, while \evetar{} is much larger, supports more IR tasks, and uniquely focuses on Arabic, it is comparable in multiple aspects to the TREC tweet test collections. This might shed some light on the expected quality of \evetar{}. 

\begin{table}[h]
\caption{Comparison of \evetar{} against six TREC microblog test collections
\label{tab:comp:trec:topics}}
\begin{threeparttable}
\begin{tabularx}{\textwidth}{lXXXXXXX}
\hline\noalign{\smallskip}
\multicolumn{1}{l}{}& {\bf \evetar} & {\bf TREC'11} & {\bf TREC'12} & {\bf TREC'13} & {\bf TREC'14} & {\bf TREC'15} & {\bf TREC'16} \\
\noalign{\smallskip}\Xcline{2-8}{2\arrayrulewidth}\noalign{\smallskip}
 {\bf Tweets} & 355M &        16M &        16M &       243M &       243M &  43M\textsuperscript{\dag} & 45M\textsuperscript{\dag}\\
{\bf Data Lang.} &     Arabic &  Multiple &  Multiple & Multiple &  Multiple &Multiple & Multiple \\
{\bf Topic Lang.} &     Arabic &      English &      English &      English &      English &      English &      English \\
{\bf Tasks} &  AS, ED, TTG, RTS & AS  & AS, Filtering  &  AS &  AS, TTG, RTS$^*$ &  TTG$^*$, RTS &  TTG$^*$, RTS \\
 {\bf Topics} &         50 &         50 &         60 &         60 &         55 &         51 &         56 \\
{\bf Qrels Pool} &  Interactive & Shared Task & Shared Task & Shared Task & Shared Task & Shared Task & Shared Task \\
 {\bf Pool Size} &      61,946	&	60,129	&	73,073	&	71,279	&	57,985	&	94,066	&	67,525 \\
 {\bf Relevant} &      24,086 &       2,965 &       6,286 &       9,011 &      10,645 &       8,233 &       3,339 \\
{\bf Rel/Topic} &      482 &       61 &      107 &      150 &      194 &      161 &       60 \\
{\bf \%Rel/Coll.} & 0.007\%  &  0.019\% & 0.039\% & 0.004\% & 0.004\%  & 0.019\%  & 0.007\%  \\
\noalign{\smallskip}\hline
\end{tabularx}
\begin{tablenotes}\footnotesize
\item[*] Task was not officially run in the track but can be supported by the test collection. 
\item[\dag] Dataset was crawled over 10 days at each participant end. Number is based on our local crawl.
\end{tablenotes}
\end{threeparttable}
\end{table}

\section{Conclusion and Future Work}
\label{sec:conclusion} 
Test collections are evaluation tools that are essential for advancing the state-of-the-art in the field of information retrieval; however, the scarcity of Arabic test collections is evident. 
In this article, we addressed the problem of creating a high-quality multi-task tweet test collection without the need of a shared-task campaign. We proposed a new language-independent approach, applied on Arabic tweets, that leverages significant events as the core of the collection design. Our study introduced \evetar{}, the first publicly-available multi-task Arabic tweet test collection. \evetar{} supports four different tasks, namely event detection, ad-hoc search, timeline generation, and real-time summarization, over Arabic tweets. 
The collection consists of 355M Arabic tweets (crawled in January 2015), 50 topics (representing significant events), 62K relevance judgments (collected via carefully-designed crowdsourcing tasks), and novelty annotations (gathered by in-house annotators). 
In addition to the full collection, we provided 4 different subsets in order to make it more accessible and further help research on dialectal Arabic IR.

We demonstrated the quality of \evetar{} from different aspects. 
First, the relevance judgments among assessors have substantial agreement (average Kappa score of 0.71), suggesting that we managed to collect high-quality judgments via crowdsourcing. Second, we evaluated a number of IR systems for each supported task and found a high correlation between the performance of systems over the full and sampled versions. Therefore, researchers who do not have resources to experiment with the full collection can also use the sampled version. Moreover, the presented performance results exhibit reference points for future studies.
Finally, \evetar{} is comparable to TREC tweet test collections in many aspects including reliability of ranking the tested systems.

There are several possible directions for future work. First, the introduction of \evetar{} opens the door for building Arabic-specific retrieval systems on Twitter. To further encourage the employment of \evetar{} and advance the research on Arabic IR, we plan to organize a shared-task evaluation campaign around the tasks supported by \evetar{}, where the performance results presented in this work would serve as strong baselines.
Second, with the success of the evaluation-as-a-service paradigm~\citep{paik2016retrievability}, we are planning to develop a search API to allow direct access to the full collection without the need of re-crawling. Finally, we plan to extend the number of labeled dialectal tweets and identify their exact dialects (e.g. Yemeni) in order to enable the evaluation of IR systems working on dialect-related tasks such as cross-dialect search.

\begin{acknowledgements}
This work was made possible by NPRP grant\# NPRP 7-1313-1-245 from the Qatar National Research Fund (a member of Qatar Foundation). The statements made herein are solely the responsibility of the authors. We would like to thank the crowd workers and in-house annotators for their valuable efforts in producing high-quality judgments of tweets.
\end{acknowledgements}
\bibliographystyle{spbasic}
\bibliography{ref}
\end{document}